\documentclass[12pt,a4paper]{article}
\input{epsf}
\global\arraycolsep=2pt %reduces the separation in eqnarrays
\begin{document}
\thispagestyle{empty}
\rightline{TTP 01-09}
\rightline{hep-ph/0103310}
\rightline{March 2001}
\bigskip
\boldmath
\begin{center}
{\bf \Large Theoretical  Aspects  of $b \to s \gamma$ Transitions }
\end{center}
\unboldmath
\smallskip
\begin{center}
{\large{\sc Thomas Mannel}}
\vspace*{2cm} \\
{\sl CERN Theory Division, CH--1211 Geneva 23, Switzerland}
\\ and \\
{\sl Institut f\"{u}r Theoretische Teilchenphysik, \\
Universit\"{a}t Karlsruhe,  D--76128 Karlsruhe, Germany}
\vspace*{1cm} \\
{\it Talk presented at the BCP4 Conference on $B$ Physics and CP
     Violation, \\ Feb. 18-23, 2001, Ise Shima, Japan} 

\end{center}
\begin{abstract}
\noindent
In this talk some of the theoretical
aspects of $b \to s \gamma$ transitions are discussed. The focus is
on inclusive decays, since these can be computed more reliably. Topics
covered are (1) perturbative QCD corrections, (2) non-perturbative
contributions and (3) effects of ``new physics'' in these decays.
\end{abstract}
\newpage
\section{Introduction}
Radiative rare $B$ decays have attracted considerable attention in the
last few years. After the first observation in 1994, by the CLEO collaboration
\cite{Alam:1995aw}, data have become quite precise \cite{Thorndike}
so that even a measurement
of the CP asymmetry in these decays \cite{Lyon,Coan:2000pu}
became possible. As far as data are
concerned, the situation clearly will improve further, after the excellent
start of both $B$ factories at KEK and at SLAC.

$B \to X_s \gamma$ tests the Standard Model (SM) in a particular way.
Since there are no tree-level contributions to these
processes in the SM, they can occur only at the
one-loop level. The GIM cancellation, which is present in all the FCNC
processes, is lifted in this case by the large top-quark mass; if
the top quark were as light as the $b$ quark, these decays would be
too rare to be observable.

Since the SM contribution is small, these decays have a good sensitivity
to ``new physics'', e.g. to new (heavy) particles contributing to the
loop. In fact, already the first CLEO data could constrain some models
for ``new physics'' in a stringent way\cite{Alam:1995aw}.

The most general effective Hamiltonian describing these decays
is given by
\begin{equation} \label{Heff}
H_{eff} = \sum_i c_i \, O_i  \, ,
\end{equation}
where the $O_i$ are local operators  
\begin{eqnarray} \label{ops}
O_{1 \cdots 6} &=& \mbox{four-fermion operators} \nonumber \\
O_7 &=& m_b  \bar{s} \sigma_{\mu \nu} (1+\gamma_5) b \, F^{\mu \nu}
\nonumber \\
O_7^\prime &=& m_s \bar{s} \sigma_{\mu \nu} (1-\gamma_5) b \,
               F^{\mu \nu}
\nonumber \\
O_8 &=& m_b \bar{s} \sigma^{\mu \nu}T^a (1+\gamma_5) b \,
       G_{\mu \nu}^a
\nonumber \\
O_8^\prime &=& m_s \bar{s} \sigma^{\mu \nu} T^a  (1-\gamma_5) b \,
    G_{\mu \nu}^a
\end{eqnarray}
and $c_i$ are pertubatively calculable coefficients. 

In any new physics analysis of $B$ decays only the coefficients $c_i$  
are tested \cite{Ali:1995bf}. 
The decay $B \to X_s \gamma$ (and the corresponding exclusive
decays) are practically determined by the two operators $O_7$ and
$O_7^\prime$, and hence these decays are mainly testing $c_7$ and
$c_7^\prime$. In the SM these two coefficients are
\begin{eqnarray}
c_7 = - \frac{G_F^2 e}{32 \sqrt{2} \pi^2}
       V_{tb} V_{ts}^* C_7  m_b, \\
c_7^\prime = - \frac{G_F^2 e}{32 \sqrt{2} \pi^2}
       V_{tb} V_{ts}^* C_7  m_s
\end{eqnarray}
where $C_7$ is a function of $(m_t/M_W)^2$, which we shall discuss later. 

Furthermore, the two
operators differ by the handedness of the quarks; in order to disentangle
these two contributions there has to be a handle on the polarization
of the quarks or of the photon, which is impossile at a $B$ factory.
Consequently, from $b \to s \gamma$ alone only the
combination $|c_7|^2 + |c_7^\prime|^2$ can be determined
in the near future.

Once the effective interaction for the quark transition is fixed, one
has to calculate from this the actual hadronic process. This step is
only for the inclusive decays under reasonable theoretical control;
for exclusive decays, form factors are needed, which either need to be
modelled or will finally come from the lattice.

For inclusive decays the machinery used is the heavy mass
expansion\footnote{A non-exhaustive selection of revies is
\cite{Manohar:2000dt,Isgur:ed.xa,Neubert:1994mb,Bigi:1997fj,Mannel:1997ky}.}.
Using this framework for the total rate
one can establish that
(1) the leading term as $m_b \to \infty$ is the free quark decay,
(2) there are no subleading corrections of order $1/m_b$,
(2) the first non-vanising corrections are of order $1/m_b^2$
and are given in terms of two parameters. This will be discussed
in section~\ref{sec:nonpert}. Additional non-perturbative uncertainties
are induced by a cut on the photon energy, which is necessary from
the exprimental point of view to suppress backgrounds. 

Part I deals with the perturbative corrections,
part II with the non-perturbative ones. In part III,
``new physics'' in $b \to s \gamma$ is considered.  

\section{PART I: Perturbative Corrections} \label{sec:pert}
The main perturbative corrections are the QCD corrections, which are
substantial. These corrections are calculated
using an effective field-theory framework.
To set this up, we have to write down first the relevant effective
Hamiltonian as in (\ref{Heff}). The operators appearing in (\ref{Heff})
mix under renormalization as we evolve down from the $M_W$ mass
scale to the relevant scale, which is the mass of the $b$ quark. 
The cofficient functions are calculated at the scale
$\mu = M_W$ as a power series in the strong coupling
\begin{equation}
c_i (M_W)= c_i^{(0)} (M_W) + \frac{\alpha_s (M_W)}{\pi} c_i^{(1)} (M_W)
+ \cdots
\end{equation}
Changing the scale $\mu$ results in a change of the coefficient functions
and in the matrix elements, such that the matrix element of the effective
Hamiltonian remains $\mu$-independent.
This change can be computed perturbatively for sufficiently large $\mu$,
using the standard machinery of renormalization
group, which involves a calculation of the anomalous-dimension
matrix that describes the mixing of the operators (\ref{ops}).

The solution of the renormalizaton group equation yields the coefficient
functions at some lower scale $\mu$, which take the form 
(schematically)
\begin{eqnarray} \label{cimu}
&&    c_i (\mu) = c_i^{(0)} (M_W)
      \sum_{n=0} b_n^{(0)}
      \left(\frac{\alpha_s}{\pi} \ln\left(\frac{M_W^2}{\mu^2}\right)\right)^n
\\ \nonumber 
&& \quad + \frac{\alpha_s}{\pi} c_i^{(1)} (M_W)
      \sum_{n=0} b_n^{(1)}
      \left(\frac{\alpha_s}{\pi} \ln\left(\frac{M_W^2}{\mu^2}\right)\right)^n
+ \cdots 
\end{eqnarray}
where the $b_n$ are obtained from the solution of the renormalization
group equation.

The last step is to compute the matrix elements of the operators at a scale
$\mu \approx m_b$. This can be done for the inclusive case using the 
$1/m_b$ expansion.
For the exclusive case, one would need the form factor  in the
corresponding approximation, which cannot be done with present
theoretical techniques.

At present, the leading and the subleading terms of the coefficients
have been calculated 
\cite{Chetyrkin:1997vx,Ali:1991tj,Adel:1994ah,Greub:1996tg},
including electroweak contributions\cite{Czarnecki:1998tn}, the
main part of which is due to the correct scale setting in $\alpha_{em}$.

A complete and up-to-date compilation can be found in
\cite{Kagan:1999ym}. Without going into any more detail we only
quote the result from \cite{Kagan:1999ym}
\begin{equation} \label{NKresult}
Br (B \to X_s \gamma) = (3.29 \pm 0.33) \times 10^{-4} \, .
\end{equation}
where this result includes a cut on the photon energy at
$E_{\gamma,min} = 0.05 \, m_b$.

The QCD corrections are in fact dramatic; they increase the rate
for $b \to s \gamma$ by about a factor of two. 
For example, already at the leading-log level we have 
$ c_7 (m_b) / c_7 (M_W) = 1.63$. Another indication
of this fact is a substantial dependence of the leading-order result on
the choice of the renormalization scale $\mu$. This is usually estimated 
by varying the scale $\mu$ between $m_b / 2$ and $2 m_b$. In this way one
obtains a variation of $\delta_\mu = {}^{+27.4\%}_{-20.4\%}$ for the
leading-order result.

Taking into account the subleading terms reduces the scale dependence
substantially. In fact, one has at subleading order  \cite{Kagan:1999ym}
$\delta_\mu = {}^{+0.1\%}_{-3.2\%}$, which is smaller than 
naively expected \cite{Buras:1994xp}. It has been argued that the
smallness of $\delta_\mu$ is accidental \cite{Kagan:1999ym}.
However, arguments have been given recently \cite{Misiak}
that these cancellations are not accidental. In fact, most of the
large radiative corrections may be assigned to the running of the
$b$ quark mass appearing in the operator $O_7$.

\section{PART II: Non-Perturbative Corrections} \label{sec:nonpert}
Non-perturbative corrections arise from different sources. 
We shall consider here
\begin{itemize}
\item Long-distance effects from intermediate vector mesons
      $ B \to J/\Psi X_s \to X_s \gamma $, 
\item Subleading terms in the heavy mass expansion:
      $1/m_b$ and $1/m_c$ corrections,  
\item Non-perturbative contributions to the  
      photon spectrum (``shape function'').
\end{itemize}

\subsection{ $ B \to J/\Psi X_s \to X_s \gamma $}

%\begin{figure}
%\centerline{\epsfysize=3truecm \epsfbox{bsglong.eps} }
%\caption{Long distance contribution
%         $ B \to J/\Psi X_s \to X_s \gamma $}
%\label{fig:bsgld}
%\end{figure}
One long-distance contribution comes from the process $B \to X_s J/\Psi$
and the subsequent decay of the (off-shell) $J/\Psi$ into a photon.
%see
% fig.~\ref{fig:bsgld}. 
The first process $B \to X_s J/\Psi$ has a branching ratio of order
1\%, at least for an on-shell $J/\Psi$. Assuming that this is
similar for the off-shell case, we have to multiply it with another factor
$1/m_c^2$ for the propagation of the $J/\Psi$ and a factor
$f_{J/\Psi}^2$, since the $J/\Psi$ has to annihilate into a photon.  
This leads us to the conclusion that this contribution is indeed
negligibly small. However, one has to keep in mind that some
extrapolation from $q^2 = m_{J/\Psi}^2$ to $q^2 = 0$ is involved,
assuming that this will not lead to a strong enhancement. 

\subsection{$1/m_b$ and $1/m_c$ corrections} 

A set of ``standard'' non-perturbative corrections arises from the
heavy mass expansion
\cite{Manohar:2000dt,Isgur:ed.xa,Neubert:1994mb,Bigi:1997fj,Mannel:1997ky}. 
As far as the total rate is concerned, we have
the subleading corrections of order $1/m_b^2$, which are parametrized in
terms of the kinetic energy $\lambda_1$  and the chromomagnetic moment
$\lambda_2$ defined by the matrix elements
\begin{eqnarray}
2 M_H \lambda_1 &=& \langle H (v) | \bar{h}_v  (iD)^2  h_v | H (v) \rangle
\\                   
6 M_H \lambda_2  &=& \langle H (v) | \bar{h}_v 
\sigma_{\mu \nu} iD^\mu iD^\nu h_v | H (v) \rangle \, .
\end{eqnarray}
In terms of these two matrix elements the total rate reads at tree level
up to order $1/m_b^2$ 
\begin{equation}
\Gamma =  \frac{G_F^2 \alpha m_b^5}{32 \pi^4} |V_{ts} V_{tb^*}|^2 |C_7|^2
\left(1 +  \frac{\lambda_1 - 9 \lambda_2}{2m_b^2} + \cdots \right) \, .
\end{equation}
This result is fully integrated over the photon
energy spectrum. One can also compute the energy spectrum of the photon
within the $1/m_b$ expansion, which is given, again at tree level, by 
\begin{eqnarray} \label{spec}
&& \frac{d\Gamma}{dx} = \frac{G_F^2 \alpha m_b^5}{32 \pi^4} 
         |V_{ts} V_{tb^*}|^2 |C_7|^2
\\
\nonumber 
&& \left(\delta(1-x) - \frac{\lambda_1 + 3 \lambda_2}{2m_b^2}
         \delta'(1-x) + \frac{\lambda_1}{6m_b^2}
         \delta '' (1-x) + \cdots \right)
\end{eqnarray}
which can only be interpreted in terms of moments of the spectrum.
We shall return to this point in the next subsection.

\begin{figure}
\centerline{\epsfysize=7truecm \epsfbox{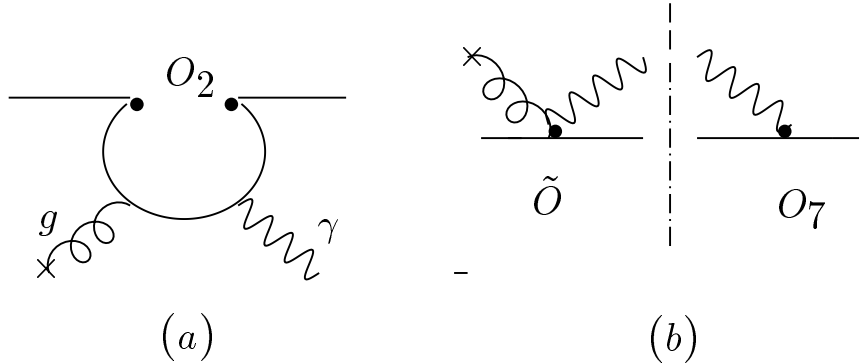} }
\vspace*{-5mm}
\caption{Interference between $O_7$ and one of the
         four-fermion operators ($O_2$),
         leading to a contribution of order $1/m_c^2$ }
\label{fig:volo}
\end{figure}

If the charm quark is also assumed to be heavy, one may discuss
the charm-mass dependence in terms of a $1/m_c$ expansion
\cite{Voloshin:1997gw}. 
The relevant contribution originates from the four-fermion operators
(e.g. the operator $O_2$) involving the charm quark,
see fig.~\ref{fig:volo}. Expanding the matrix element of $O_2$ in powers
of $1/m_c$ we obtain a local operator of the form 
\begin{equation}
O_{1/m_c^2} = \frac{1}{m_c^2} \bar{s} \gamma_\mu (1-\gamma_5) T^a b \, 
      G^a_{\nu \lambda} \epsilon^{\mu \nu \rho \sigma}
      \partial^\lambda F_{\rho \sigma}
\end{equation}
which can interfere with the leading term $O_7$ (see fig.~\ref{fig:volo}).  

The detailed calculation \cite{Buchalla:1998ky,Ligeti:1997tc}
reveals that this contribution is rather small
\begin{equation}
\frac{\delta \Gamma_{1/m_c^2}}{\Gamma} =
  -\frac{C_2}{9 C_7} \frac{\lambda_2}{m_c^2} \approx 0.03 
\end{equation}
\subsection{Non-perturbative corrections in the photon energy spectrum} 
The non-perturbative corrections for the total rate are thus quite small
and can safely be neglected against the perturbative ones. However, to
extract the process $B \to X_s \gamma$ there has to be a lower cut on the
photon energy to get rid of the uninteresting processes such as ordinary
bremsstrahlung. Clearly it is desirable to have this cut as high as
possible, but this makes the process ``less inclusive'' and hence more
sensitive to non-perturbative contributions to the photon-energy spectrum.

Since we are dealing at tree level with a two-body decay, the 
naive calculation of the photon spectrum yields a $\delta$ function
at partonic level and the $1/m_b^n$ corrections are again distributions
located at the partonic energy $E_\gamma = m_b /2$, see (\ref{spec}).  
Clearly (\ref{spec}) cannot be used to implement a cut on the photon
energy spectrum, since this is not a smooth function.

The perturbative contributions have been calculated and yield a spectrum
that is mainly determined by the bremsstrahlung of a radiated gluon.
This part of the calculation is fully perturbative and enters the
next-to-leading order analysis described in part I of this talk. In
particular, the partonic $\delta$ function smoothens and turns
into  ``plus distributions'' of the form
\begin{equation} \label{radcorr}
\frac{d\Gamma}{dx} = \cdots + \frac{\alpha_s}{\pi}
                   \left[ \left(\frac{\ln(1-x)}{1-x} \right)_+ \, , \,
                          \left(\frac{1}{1-x} \right)_+ \right] \, ,
\end{equation}
where the ellipses denote terms that are regular as $x \to 1$ and 
contributions proportional to $\delta(1-x)$, which are determined by 
virtual gluons.

Here we shall focus on the non-perturbative contributions close to the
endpoint. The general structure of the terms in the $1/m_b$ expansion
is 
\begin{eqnarray} \label{general}
\frac{d\Gamma}{dx} &=& \Gamma_0  \left[ \sum_i a_i  
                     \left(\frac{1}{m_b} \right)^i 
                     \delta^{(i)} (1-x) \right. \\ \nonumber  
&& \qquad  \left. 
                     + {\cal O} ((1/m_b)^{i+1} \delta^{(i)} (1-x))  
\right] \, ,
\end{eqnarray}
where $\delta^{(i)}$ is the $i$th derivative of the $\delta$ function. 

It has been shown \cite{Neubert:1994um,Bigi:1994ex} that the terms
with $\delta^{(i)} (1-x)/m_b^i$ can be resummed into a non-perturbative
function such that the photon energy spectrum becomes 
\begin{equation} 
\frac{d\Gamma}{dx} = \frac{G_F^2 \alpha m_b^5}{32 \pi^4} 
         | V_{ts} V_{tb^*}|^2 |C_7|^2  f(m_b[1-x]) \, , 
\end{equation}
where the non-perturbative fuction $f$ is formally defined by the
matrix element 
\begin{equation}
2M_B f(\omega) = \langle B | \bar{Q}_v \delta(\omega + iD_+) Q_v | B 
                 \rangle  \, . 
\end{equation}
Here $D_+$ is the light-cone component of the covariant derivative, acting
on $Q_v$, which denotes a heavy-quark field in the static approximation.   

The shape function is in fact a universal function, which appears 
for any heavy-to-light transition in the corresponding kinematical
region. In
general these transitions should  be written as a convolution of a
(perturbatively calculable) Wilson coefficient and the non-perturbative
matrix element
\begin{equation}
d\Gamma =  \int\! d\omega\, C_0(\omega)
\langle B | O_0(\omega)  | B \rangle  
\end{equation}
with
\begin{equation}
O_0 (\omega) =  \bar{Q}_v \delta(\omega + iD_+) Q_v
\end{equation}
At tree level this leads to a simple and intuitive formula in which
the mass $m_b$ is replaced by an ``effective mass''
$m_b^* = m_b - \omega$ such that
\begin{equation} \label{masscon}
d\Gamma =  \int\! d\omega\, d\Gamma_{tree} (m_b \to m_b^*) f(\omega)
\end{equation}

Since this function is universal, it appears in the semileptonic
$b \to u \ell \bar{\nu}$ transitions as well as in the $b \to s \gamma$
decays. At leading twist, this leads to a model-independent relation
between these inclusive decays, which may be used to obtain
$| V_{ts}/V_{ub} |$ \cite{Mannel:1999gs,Leibovich:2000ey}.

Moments of the shape function can be related to the parameters describing
the subleading effects in the $1/m_b$ expansion. One has
\begin{eqnarray}
&& \int d\omega \, f(\omega) = 1 \, ,\quad
\int d\omega \, \omega \, f(\omega) = 0 \, , \\ \nonumber 
&& \int d\omega \, \omega^2 \, f(\omega) = - \frac{\lambda_1}{3 m_b^2}
\, , \quad \int d\omega \, \omega^2  \, f(\omega)
       = - \frac{\rho_1}{3 m_b^2} \, .
\end{eqnarray}

Radiative corrections can be included using 
the machinery of effective field theory as described in part~I.
However, here some ambiguity arises from the 
appearance of a double logarithm (see (\ref{radcorr})),
which makes the matching ambiguous. Various authors
\cite{Mannel:1999gs,DeFazio:1999sv} have used the mass convolution
formula (\ref{masscon}), although this has not yet been proven to be
correct beyond the tree level. 

Finally one may also try to resum the subleading terms in $1/m_b$, i.e.
the terms of order $\delta^{(i)} (1-x)/m_b^{i+1}$ in (\ref{general}).  
This has been discussed in \cite{Bauer:2001mh},
where it has been shown that the relevant operators are 
\begin{eqnarray}
&& O_1^\mu(\omega) =  
\bar Q_v \left\{i D^\mu,\delta(iD_+ + \omega)\right\} Q_v \\
&& O_2^\mu(\omega) = \nonumber 
i\bar Q_v \left[i D^\mu,\delta(i D_+ + \omega)\right]  Q_v \\
&& O_3^{\mu\nu}(\omega_1,\omega_2)  = \\ \nonumber  
&& \quad \bar Q_v \delta(iD_+ +\omega_2) \left\{iD^\mu_\perp, 
iD^\nu_\perp\right\}\delta(iD_+ +\omega_1)  Q_v  \nonumber \\
&& \nonumber O_4^{\mu\nu}(\omega_1,\omega_2) = \\ \nonumber 
&& \quad g_s \bar Q_v \delta(iD_+ + \omega_2) G_\perp^{\mu\nu}
  \delta(iD_+ +\omega_1) Q_v  \nonumber \, , 
\end{eqnarray}
plus the corresponding ones where a Pauli spin matrix appears between
the quark spinors. 

\begin{figure}
\centerline{\epsfysize=7truecm \epsfbox{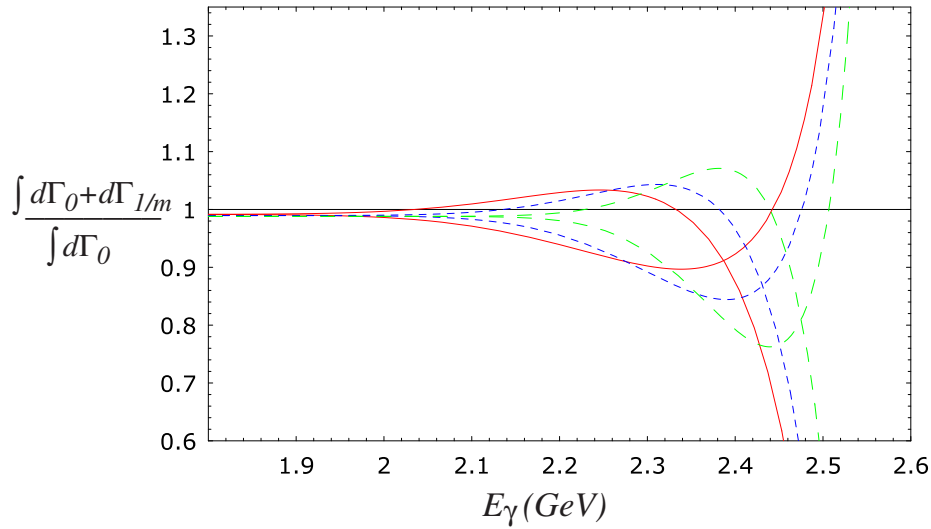} }
\caption[]{{\small Partially integrated rate 
           normalized to the leading twist result. 
The three lines with a peak correspond to $\rho_2 = (500 \mbox{ MeV})^3$, 
and $\bar\Lambda = 570$ MeV (solid line),  
$\bar\Lambda = 470$ MeV (short-dashed line) 
$\bar\Lambda = 370$ MeV (long-dashed line). The two lines with a dip
have $\rho_2 = -(500 \mbox{ MeV})^3$ and $\bar\Lambda = 470$ MeV
(dashed line),  
$\bar\Lambda = 370$ MeV (dotted line).}} 
\label{sublead}
\end{figure}

The effect of the subleading terms can be parametrized by four universal
functions, which appear again in both $b \to u \ell \bar{\nu}$
and $b \to s \gamma$. Using a simple but realistic model
the effects of the subleading terms may be estimated
as a function of the lower photon energy
cut. In fig.~\ref{sublead} we plot the rate integrated from a lower cut
as a function of this cut for various values of the parameters.
As expected, the subleading terms at cut values of 2.3 GeV are of order
10\% and negligibly small below 2.1 GeV.

\boldmath
\section{PART III: ``New Physics'' in $b \to s \gamma$} \label{sec:newphys}
\unboldmath
In the Standard Model, $b \to s \gamma$ is a loop-induced process; it
thus has considerable sensitivity to new physics effects. 
However, as already pointed out in the introduction, 
any $B$ physics experiment tests the coefficients $c_i$ appearing  
in the effective Hamiltonian (\ref{Heff}) and thus 
all the information on new effects is encoded in combinations of 
the low-energy parameters $c_i$,
which have to be computed in the Standard Model with
the best possible accuracy. Comparing this to $B$ decay data, it 
will clearly be impossible to find clean evidence for
some specific scenario of new physics. 

At present, no significant deviation from the Standard Model has
been observed in $B \to X_s \gamma$ nor in any other $B$ decay.  
Given that there are processes that are sensitive to new effects,
$B$ physics (and $b \to s \gamma$ in particular) can contribute
to constrain new physics scenarios.

Keeping this in mind one may try various scenarios of new physics
and calculate the effects on $b \to s \gamma$, i.e.\ calculate the
coefficients of the low energy effective Hamiltonian 
(\ref{Heff}) in specific scenarios.
There is an enormous variety of models for physics beyond
the Standard Model on the market, and is is impossible 
to cover all these ideas.

For that reason I shall only consider two examples, which
are instructive and demonstrate the kind of sensitivity one may expect.
In the next subsection I shall consider the Type-II two-Higgs doublet model 
and in subsection~\ref{sec:susy} I shall discuss a few recent papers on
supersymmetry with large values of $\tan \beta$.

\subsection{Two-Higgs-Doublet Model (Type II)}
One popular and consistent way to extend the Standard Model is to add
one or more Higgs doublets. This can be done in various ways, but one
well motivated way is to have two Higgs doublets where one doublet gives
the mass to the up quarks, the other doublet to the down quarks.

\begin{figure}
\centerline{\epsfysize=7truecm \epsfbox{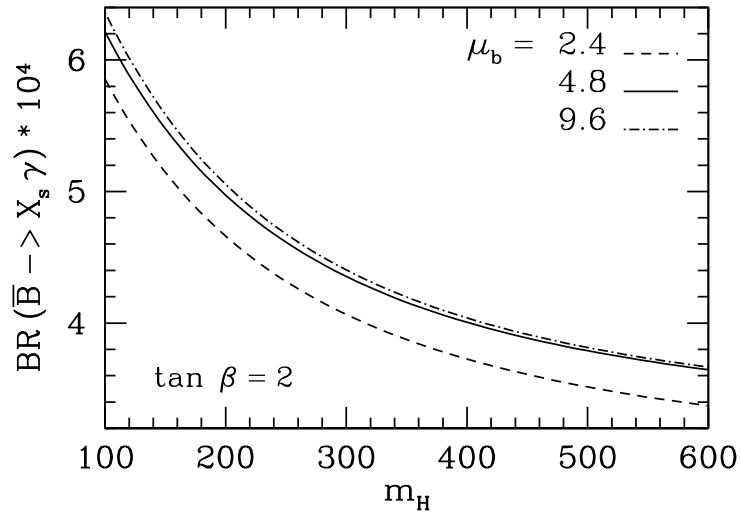} }
\caption[]{The branching ratio for $B \to X_s \gamma$ as a function of the
         charged Higgs mass; figure taken from
         \protect{\cite{Borzumati:1998tg}}.}
\label{2hdmfig1}
\end{figure}

\begin{figure}
\centerline{\epsfysize=7truecm \epsfbox{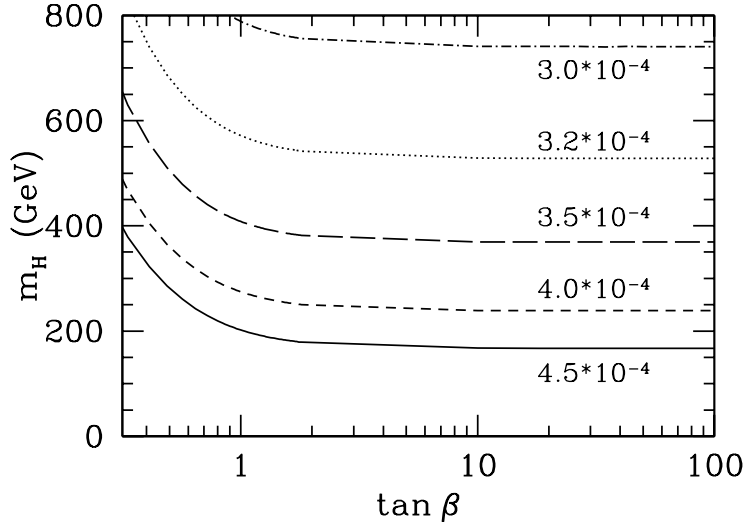} }
\caption[]{The $\tan \beta$-$M_{H^+}$ plane; contours indicate different
         experimental values for $B \to X_s \gamma$;
         figure taken from
         \protect{\cite{Borzumati:1998tg}}.} 
\label{2hdmfig2}
\end{figure}

Out of the eight degrees of freedom of the Higgs sector,
three are needed to give mass
to the heavy weak bosons, while the other five become physical states.  
In particular, the spectrum contains a charged Higgs boson, which
appears in the loops relevant to $b \to s \gamma$. The first analysis of
this decay in this type of two-Higgs doublett model has been performed in
\cite{Ciuchini:1998xe}.

The parameters of this model are the ratio of the two vacuum expectation 
values (usually expressed as $\tan \beta = v_1/v_2$), the mass $M_{H^+}$
of the charged Higgs boson, all other parameters are irrelevant for our
discussion.

In fig.~\ref{2hdmfig1} (taken from \cite{Borzumati:1998tg})
the branching ratio
of $b \to s \gamma$ is plotted as a function of the charged Higgs mass, for
three different values of the renormalization scale $\mu$.

In fig.~\ref{2hdmfig2} (taken from \cite{Borzumati:1998tg})  
we plot contours in the $\tan \beta$--$M_{H^+}$ plane for different values
of the $B \to X_s \gamma$ branching ratio. From this figure it becomes
clear that there is no large effect induced by enlarging $\tan \beta$. 
One may derive bounds on the charged Higgs mass independently of
$\tan \beta$; the current bound is
$M_{H^+} > 314$ GeV at 95\% CL \cite{Gambino}.
This does not yet include the new CLEO result \cite{Thorndike},
which will move the bound to even higher values.

\subsection{Supersymmetry with large $\tan \beta$}  \label{sec:susy}

\begin{figure}
\centerline{\epsfysize=7truecm \epsfbox{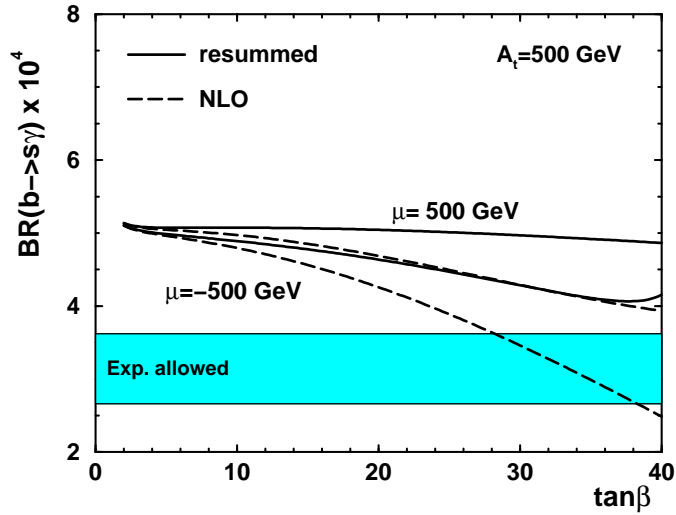} }
\caption[]{The rate for $B \to X_s \gamma$ versus $\tan \beta$; for the
values of the parameters see text. Figure taken from
\protect{\cite{Carena:2001uj}}.}
\label{susy1}
\end{figure}

If supersymmetry were an exact symmetry, $b \to s \gamma$ would vanish,
owing to the cancellations between particles and sparticles
\cite{Ferrara:1974wb}. This
means that $b \to s \gamma$ tests the breaking of supersymmetry.
Clearly many different scenarios for this symmetry breaking can be
invented, having complicated flavour structure.

Again I shall pick an example from a recent analysis
\cite{Degrassi:2000qf,Carena:2001uj}.
In these papers it has been pointed out that $B \to X_s \gamma$ can indeed
be enhanced in scenarios whith large $\tan \beta$. Working in the MSSM
with a flavour-diagonal supersymmetry-breaking sector, the relevant
parameters are the charged Higgs mass $M_{H^+}$, the light stop mass
$m_{\tilde{t}_1}$, the supersymmetric $\mu$ parameter, and the parameter
$A_t$ from the sector of soft-supersymmetry breaking. 

For large $\tan \beta$, renormalization group methods may be used to
resum these terms \cite{Carena:2001uj} and one may confront these results with
the recent data. In fig.~\ref{susy1} (taken from \cite{Carena:2001uj})
we plot the
rate for $B \to X_s \gamma$ as a function of $\tan \beta$ for
$\mu = \pm 500$ GeV; the values of the other parameters are
$M_{H^+} = 200$ GeV, $m_{\tilde{t}_1} = 250$ GeV, all other
sparticle masses being at 800 GeV. 

A similar plot can be made for negative $A_t$, but for the parameters
chosen here this scenario is already practically excluded.

\begin{figure}
\centerline{\epsfysize=7truecm \epsfbox{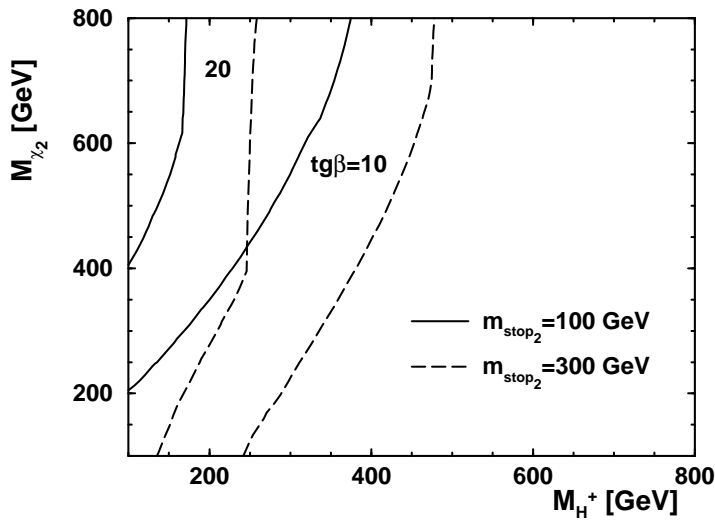} }
\caption[]{Allowed range in the $M_{H^+}$--$M_{\tilde{\chi}_2}$ plane
for different values of $\tan \beta$ for two values of $m_{\tilde{t}_2}$.
Figure taken from \protect{\cite{Carena:2001uj}}.}
\label{susy3}
\end{figure}

Given such a scenario, one may also scan over some range for the
parameters and identify regions that are still allowed by the
experimental constraints. In fig.~\ref{susy3} such a scan was performed
with $m_{\tilde{t}_2} \le m_{\tilde{t}_1} \le 1$ TeV,
$m_{\tilde{\chi}_2^+} \le m_{\tilde{\chi}_1^+} \le 1$ TeV,
$|A_t| \le 500$ GeV,  all other sparticle masses being 1 TeV.

Clearly $B \to X_s \gamma$ places significant constraints on the parameter
space of certain supersymmetric scenarios; however, these studies have been
performed with a flavour diagonal supersymmetry-breaking sector. An analysis
witout this constraint can be found in \cite{Gabbiani:1996hi}

\section{Conclusion}
The inclusive radiative rare decay $B \to X_s \gamma$ is under reasonable
theoretical control; the latest theoretical prediction \cite{Misiak} is
slightly higher than (\ref{NKresult})
$$
Br_{th} (B \to X_s \gamma) = (3.71 \pm 0.30) \times 10^{-4} \, ,
$$
where the difference originates from a different value for the  
ratio $m_c / m_b$; while \cite {Kagan:1999ym} use the ratio of
pole masses, in \cite{Misiak} $m_c^{\overline{MS}}/m_b^{Pole}$ is
is used as an ``educated guess'' of NNLO corrections. 

The latest (combined) experimental result is \cite{Misiak} 
\begin{equation}
Br_{exp} = (B \to X_s \gamma) = (2.96 \pm 0.35) \times 10^{-4} \, ,
\end{equation}
which is in agreement with theory within $1.6\sigma$.   

The theoretical uncertainty is mainly determined by our ignorance of
some of the input parameters (quark masses, mixing angles) and to some
extent also by the uncertainty of higher-order radiative corrections. 
Improving the current theoretical uncertainty will be very difficult
with current theoretical tools. 

\section*{Acknowledgements}
I thank Toni Sanda for organizing such an exciting meeting in such a
beautiful place, Tobias Hurth for comments on the manuscript, and
Paolo Gambino for useful discussions.

\end{document}